\documentclass[twocolumn,showpacs,preprintnumbers,amsmath,amssymb,prl]{revtex4}

\usepackage{graphicx}
\usepackage{color}
\usepackage{amsmath}
\usepackage{amssymb}
\usepackage{exscale}
\usepackage{anysize}
\marginsize{1.5cm}{1.5cm}{0.7cm}{1cm}

\newcommand{\ee}{\textrm{e}}
\newcommand{\ii}{\mathrm{i}}
\newcommand{\dd}{\mathrm{d}}
\newcommand{\qq}{\mathbf{q}}

\newcommand{\br}{\mathbf{r}}

\renewcommand{\Re}{\mathfrak{Re}}

\newcommand{\an}[1]{\hat{#1}}
\newcommand{\cre}[1]{\hat{#1}^\dag}

\begin{document}

\title{Polaron physics in optical lattices}

\author{Martin Bruderer, Alexander Klein, Stephen R. Clark, and Dieter Jaksch}

\affiliation{Clarendon Laboratory, University of Oxford, Parks
Road, Oxford OX1 3PU, United Kingdom}

\begin{abstract}
We investigate the effects of a nearly uniform Bose-Einstein
condensate (BEC) on the properties of immersed trapped impurity
atoms. Using a weak-coupling expansion in the BEC-impurity
interaction strength, we derive a model describing polarons, i.e.,
impurities dressed by a coherent state of Bogoliubov phonons, and
apply it to ultracold bosonic atoms in an optical lattice. We show
that, with increasing BEC temperature, the transport properties of
the impurities change from coherent to diffusive. Furthermore,
stable polaron clusters are formed via a phonon-mediated off-site
attraction.
\end{abstract}

\date{\today}
\pacs{03.75.-b, 03.67.-a, 71.38.Mx, 71.38.Ht}

\maketitle

% 03.67.-a    Quantum information
% 03.75.-b    Matter waves
% 03.75.Ss    Degenerate Fermi gases
% 71.38.-k    Polarons and electron-phonon interactions
% 71.38.Fp    Large or Froehlich polarons
% 71.38.Ht    Self-trapped or small polarons
% 71.38.Mx    Bipolarons
% 74.20.Mn    Nonconventional mechanisms (spin fluctuations, polarons and bipolarons, resonating valence bond model, anyon mechanism, marginal Fermi liquid, Luttinger liquid, etc.) (Superconductivity)

\maketitle

%%%%%%%%%%%%%%%%%%%%%%%%%%%%%%%%%%%%%%%%%%%%%%%%%%%%%%%%%%%%%%%%%%%%%%%%%
%%%%%%%%%%%%%%%%%%%%%%%%%%%%%%%%%%%%%%%%%%%%%%%%%%%%%%%%%%%%%%%%%%%%%%%%%

The lack of lattice phonons is a distinguishing feature of optical
lattices, i.e.,~conservative optical potentials formed by
counterpropagating laser beams, and contributes to the excellent
coherence properties of atoms trapped in them
\cite{Bloch-NatPhys-2005}. However, some of the most interesting
phenomena in condensed matter physics involve phonons, and thus it
is also desirable to introduce them in a controlled way into
optical lattices. Recently, it has been shown that immersing an
optical lattice into a Bose-Einstein condensate (BEC) leads to
interband phonons, which can be used to load and cool atoms to
extremely low temperatures \cite{Griessner-PRL-2006}. Here, we
instead concentrate on the dynamics within the lowest Bloch band
of an immersed lattice, and show how intraband phonons lead to the
formation of polarons \cite{Alexandrov-Mott-1994,Mahan-2000}. This
has a profound effect on lattice transport properties, inducing a
crossover from coherent to incoherent hopping as the BEC
temperature increases. Furthermore, polarons aggregate on {\em
adjacent} lattice sites into stable clusters, which are not prone
to loss from inelastic collisions. Since these phenomena are
relevant to the physics of conduction in solids, introducing
phonons into an optical lattice system may lead to a better
understanding of high-temperature superconductivity
\cite{Alexandrov-PRB-1986,Alexandrov-Mott-1994} and charge
transport in organic molecules \cite{Henderson-Proc-1999}.
Additionally, this setup may allow the investigation of the
dynamics of classically indistinguishable particles
\cite{Gottesman-2005}.

\begin{figure}[t]
  \includegraphics[width=7cm]{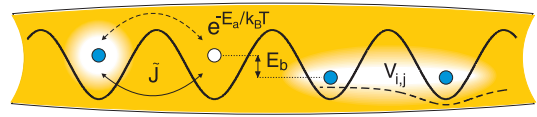}
  \caption{(Color online) A quantum degenerate gas confined to an
optical lattice is immersed in a much larger BEC. For increasing
BEC temperature $T$, a crossover from coherent to diffusive
hopping, characterized by $\tilde{J}$ and $E_a$, respectively, can
be observed. The phonon-induced interaction potential $V_{i,j}$
leads to the formation of off-site polaron clusters, separated by
a gap $E_b$ from the continuum of unbound states.
  \label{Fig:Scheme}}
\end{figure}

Experimental progress in trapping and cooling atoms has recently
made a large class of interacting many-body quantum systems
\cite{Lewenstein-2006} accessible. For instance, the formation of
repulsively bound atom pairs on a single site has been
demonstrated \cite{Winkler-nature-2006-short}, and strongly
correlated mixtures of degenerate quantum gases have been realized
\cite{ref-exp-fermi-bose-short}. In such Bose-Fermi mixtures, rich
phase diagrams can be expected, including charge and spin density
wave phases \cite{Mathey-PRL-2004,Pazy-PRA-2005}, pairing of
fermions with bosons \cite{Lewenstein-PRL-2004}, and a supersolid
phase \cite{Buchler-PRL-2003}.  Here we instead consider one
atomic species, denoted as the impurities, confined to a trapping
potential, for example an optical lattice, immersed in a nearly
uniform BEC, as shown in Fig.~\ref{Fig:Scheme}. Based on a
weak-coupling expansion in the BEC-impurity interaction strength,
we derive a model in terms of polarons, which are composed of
impurity atoms dressed by a coherent state of Bogoliubov phonons
\cite{Alexandrov-Mott-1994,Mahan-2000}. The model also includes
attractive impurity-impurity interactions mediated by the phonons
\cite{Bardeen-PR-1967,Klein-PRA-2005}. An essential requirement
for our model is that neither interactions with impurities nor the
trapping potential confining the impurities impairs the ability of
the surrounding gas to sustain phononlike excitations. The first
condition limits the number of impurity atoms
\cite{ref-exp-fermi-bose-short}, whereas the latter requirement
can be met by using a species-specific optical lattice potential
\cite{Leblanc-2007}. Moreover, unlike in the case of
self-localized impurities \cite{Sacha-PRA-2006}, we assume that
the one-particle states of the impurities are not modified by the
BEC, which can be achieved by sufficiently tight impurity
trapping.

%%%%%%%%%%%%%%%%%%%%%%%%%%%%%%%%%%%%%%%%%%%%%%%%%%%%%%%%%%%%%%%%%%%%
%%%%%%%%%%%%%%%%%%%%%%%%%%     Model    %%%%%%%%%%%%%%%%%%%%%%%%%%%%
%%%%%%%%%%%%%%%%%%%%%%%%%%%%%%%%%%%%%%%%%%%%%%%%%%%%%%%%%%%%%%%%%%%%

%%%%%%%%%%%%%%%%%%%%%%%%%%%%%%%%%%%%%%%%%%%%%%%%%%%%%%%%%%%%%%%%%%%%
%%%%%%%%%%%%%%%%%%%%%%%%%%%%%%%%%%%%%%%%%%%%%%%%%%%%%%%%%%%%%%%%%%%%

\textit{Model.---}The Hamiltonian of the system is composed of
three parts, $\hat{H} = \hat{H}_{\mathrm{\chi}} +
\hat{H}_{\mathrm{B}} + \hat{H}_{\mathrm{I}}$, where $\hat H_\chi$
governs the dynamics of the impurity atoms, which can be either
bosonic of fermionic. The BEC Hamiltonian $\hat{H}_{\mathrm{B}}$
and the density-density interaction Hamiltonian
$\hat{H}_{\mathrm{I}}$ are
\begin{equation}\nonumber
    \hat{H}_{\mathrm{B}}=\int\dd\br\,\cre{\phi}(\br)\!\left[-\frac{\hbar^2\nabla^2}{2
    m_b}+V_{\mathrm{ext}}(\br)+\frac{g}{2}\cre{\phi}(\br)\an{\phi}(\br)\right]\!\an{\phi}(\br)\,,
\end{equation}
\begin{equation}\nonumber
    \hat{H}_{\mathrm{I}}=\kappa\int
    \dd\br\,\cre{\chi}(\br)\an{\chi}(\br)\cre{\phi}(\br)\an{\phi}(\br)\,,
    \nonumber
\end{equation}
where $\an{\chi}(\br)$ is the impurity field operator and
$\an{\phi}(\br)$ is the condensate atom field operator satisfying
the commutation relations
$[\an{\phi}(\br),\cre{\phi}(\br^\prime)]=\delta(\br-\br^\prime)$
and $[\an{\phi}(\br),\an{\phi}(\br^\prime)] = 0$. The coupling
constants $g>0$ and $\kappa$ account for the boson-boson and
impurity-boson interaction respectively, $m_b$ is the mass of a
condensate atom and $V_{\mathrm{ext}}(\br)$ a weak external
trapping potential. Without yet specifying $\hat H_\chi$ we expand
$\an{\chi}(\br)=\sum_\nu\eta_\nu(\br)\an{a}_\nu$, where
$\eta_\nu(\br)$ are a set of orthogonal mode functions of the
impurities and $\an{a}_\nu$ ($\cre{a}_\nu$) the corresponding
annihilation (creation) operators, labeled by the quantum numbers
$\nu$.

A common approach to find the elementary excitations of the BEC in
the presence of impurities is to solve the Gross-Pitaevskii
equation (GPE) \cite{Oehberg-PRA-1997-short} for the full system,
i.e.,~for $\kappa\neq 0$, and to subsequently quantize small
oscillations around the classical ground state. To obtain a
quantum description of the impurity dynamics, we instead solve the
GPE without taking $\hat H_I$ into account, and express the BEC
deformations around the impurities as coherent states of
Bogoliubov phonons. Specifically, we write $\an{\phi}(\br) =
\phi_0(\br) + \delta\an{\phi}(\br)$, with
$\phi_0(\br)=\phi^\ast_0(\br)$ the solution of the GPE for $\kappa
= 0$. Provided that the impurity-boson coupling is sufficiently
weak, i.e.,~$|\kappa|/g n_0(\br)\xi^D(\br)\ll1$, with $\xi(\br) =
\hbar/\sqrt{m_b g n_0(\br)}$ the healing length,
$n_0(\br)=\phi^2_0(\br)$ and $D$ the number of spatial dimensions,
we expect that the deviation of $\an{\phi}(\br)$ from
$\phi_0(\br)$ is of order $\kappa$,
i.e.,~$\langle\delta\an{\phi}(\br)\rangle\propto\kappa$, where
$\langle\,\cdot\,\rangle$ stands for the expectation value. We
insert $\phi_0(\br) + \delta\an{\phi}(\br)$ into the Hamiltonian
$\hat{H}_{\mathrm{B}} + \hat{H}_{\mathrm{I}}$, keep terms up to
second order in $\kappa$, and obtain the linear term $\kappa\int
    \dd\br\,\cre{\chi}(\br)\an{\chi}(\br)\phi_0(\br)[\delta\cre{\phi}(\br) +
    \delta\an{\phi}(\br)]$, in addition to the standard constant and quadratic terms in
$\delta\an{\phi}(\br)$ and $\delta\cre{\phi}(\br)$, since
$\phi_0(\br)$ is no longer the ground state of the system.

In order to diagonalize the quadratic terms in
$\delta\an{\phi}(\br)$ and $\delta\cre{\phi}(\br)$, we use the
expansion $\delta\an{\phi}(\br)=\sum_\mu[u_\mu(\br)\an{b}_\mu -
v^\ast_\mu(\br)\cre{b}_\mu]$, where $u_\mu(\br)$ and
$v^\ast_\mu(\br)$ are the solutions of the Bogoliubov--de\,Gennes
equations \cite{Oehberg-PRA-1997-short} for $\kappa = 0$, and
$\an{b}_\mu$ ($\cre{b}_\mu$) are the bosonic Bogoliubov
annihilation (creation) operators, labeled by the quantum numbers
$\mu$. We assume that the mode functions $\eta_\nu(\br)$ are
localized on a length scale much smaller than is set by
$V_{\mathrm{ext}}(\br)$, and that
$\int\dd\br\,|\eta_\nu(\br)|^2|\eta_\tau(\br)|^2\approx 0$ for
$\nu\neq\tau$, i.e.,~the probability densities $|\eta_\nu(\br)|^2$
for different mode functions deviate appreciably from zero only
within mutually exclusive spatial regions. In this case
$\int\dd\br\,\phi_0(\br)[u_\mu(\br) -
v_\mu(\br)]\eta_\nu(\br)\eta^\ast_{\tau}(\br)\approx 0$ and
$\int\dd\br\,n_0(\br)\eta_\nu(\br)\eta^\ast_{\tau}(\br)\approx 0$
hold for $\nu\neq\tau$, and hence the nondiagonal impurity-phonon
coupling is negligible. The total Hamiltonian can thus be
rewritten in the form of a Hubbard-Holstein model
\cite{Holstein-Ann-1959} $\hat{H} = \hat{H}_\chi +
\sum_{\nu,\mu}\hbar\omega_\mu(M_{\nu,\mu}\an{b}_\mu +
    M^\ast_{\nu,\mu}\cre{b}_\mu)\an{n}_\nu+\sum_\nu \bar{E}_\nu\,\hat{n}_\nu + \sum_\mu \hbar\omega_\mu\cre{b}_\mu\an{b}_\mu$,
with $\hbar\omega_\mu$ the energies of the Bogoliubov excitations,
the number operator $\hat{n}_\nu = \cre{a}_\nu\an{a}_\nu$, the
dimensionless matrix elements
$M_{\nu,\mu}=(\kappa/\hbar\omega_\mu)\int\dd\br\phi_0(\br)\left[u_\mu(\br)
- v_\mu(\br)\right]|\eta_\nu(\br)|^2$ and the mean field shift
$\bar{E}_\nu = \kappa \int\dd\br\,n_0(\br)|\eta_\nu(\br)|^2$. We
obtain an effective Hamiltonian $\hat{H}_\mathrm{eff}$ including
corrections to $\phi_0(x)$ of order $\kappa$ by applying the
unitary Lang-Firsov transformation
\cite{Alexandrov-Mott-1994,Mahan-2000}
$\hat{H}_\mathrm{eff}=\hat{U}\hat{H}\hat{U}^\dag$, with $\hat U =
\exp\big[
\sum_{\nu,\mu}(M^\ast_{\nu,\mu}\cre{b}_\mu-M_{\nu,\mu}\an{b}_\mu)\hat{n}_\nu\big]$,
which yields
\begin{eqnarray}\label{hlf}\nonumber
    \hat{H}_{\mathrm{eff}}&=&\hat{U}\hat{H}_{\chi}\hat{U}^\dag +
    \sum_\nu (\bar{E}_\nu-E_\nu)\hat{n}_\nu - \sum_\nu
    E_\nu\hat{n}_\nu(\hat{n}_\nu-1)\\
    &&-\frac{1}{2}\sum_{\nu\neq\tau}V_{\nu,\tau}\hat{n}_\nu\hat{n}_{\tau} + \sum_\mu
    \hbar\omega_\mu\cre{b}_\mu\an{b}_\mu\,.
\end{eqnarray}
The transformed impurity Hamiltonian
$\hat{U}\hat{H}_\chi\hat{U}^\dag$ is obtained using the relation
$\hat{U}\cre{a}_\nu\hat{U}^\dag =
    \cre{a}_\nu\cre{X}_\nu$, where $\cre{X}_\nu =
\exp\big[\sum_\mu(M^\ast_{\nu,\mu}\cre{b}_\mu-M_{\nu,\mu}\an{b}_\mu)\big]$
is a Glauber displacement operator that creates a coherent phonon
cloud, i.e.,~a BEC deformation, around the impurity. In the limit
where the BEC adjusts instantaneously to the impurity
configuration, polarons created by $\cre{a}_\nu\cre{X}_\nu$ are
the appropriate quasiparticles, and $\hat{H}_{\mathrm{eff}}$
describes a nonretarded interaction with the potential
$V_{\nu,\tau} = \sum_\mu\hbar\omega_\mu\left(M_{\nu,\mu}
M^\ast_{\tau,\mu} +M^\ast_{\nu,\mu} M_{\tau,\mu}\right)$. The
polaronic level shift $E_\nu = \sum_\mu\hbar\omega_\mu
|M_{\nu,\mu}|^2$ is equal to the characteristic potential energy
of an impurity in the deformed BEC.

%%%%%%%%%%%%%%%%%%%%%%%%%%%%%%%%%%%%%%%%%%%%%%%%%%%%%%%%%%%%%%%%%%%%
%%%%%%%%%%%%%%%%%%%%%%%%%%     Application    %%%%%%%%%%%%%%%%%%%%%%
%%%%%%%%%%%%%%%%%%%%%%%%%%%%%%%%%%%%%%%%%%%%%%%%%%%%%%%%%%%%%%%%%%%%

We now turn to the specific case of bosons loaded into an optical
lattice immersed in a homogeneous BEC \cite{Griessner-PRL-2006}.
In the tight-binding approximation, the impurity dynamics is well
described by the Bose-Hubbard model
 $\hat H_{\chi} =
  -J \sum_{\langle i,j \rangle} \hat a^\dagger_{i} \hat a_{j} +
  \frac{1}{2}U\sum_j \hat{n}_j(\hat{n}_j-1)
  + \mu\sum_{j}\hat n_j \,$, where $\mu$ describes the energy
  offset, $U$ the on-site interaction strength, and $J$ the
  hopping matrix element between adjacent sites \cite{Lewenstein-2006,ref-lattice}.
The modes of the lattice atoms are Wannier functions $\eta_j(\br)$
of the lowest Bloch band localized at site $j$, and $\langle
i,j\rangle$ denotes the sum over nearest neighbors. Noting that
$[\hat{U},\hat{n}_j]=0$, we find
\begin{equation}\label{htrans}
\hat{U}\hat{H}_\chi\hat{U}^\dag=-J\mbox{$\sum\limits_{\langle i,j
\rangle}$}(\hat X_i\hat a_i)^\dagger \hat X_j\hat a_{j}
+\frac{U}{2}\mbox{$\sum\limits_{j}$}\hat{n}_j(\hat{n}_j-1)+
    \mu\mbox{$\sum\limits_{j}$}\hat{n}_j\,,
\end{equation}
with the corresponding matrix elements $M_{j,\qq} = \kappa
\sqrt{n_0 \varepsilon_{\qq}/(\hbar\omega_{\qq})^3} \, f_j(\qq)$,
where $\qq$ is the phonon momentum, $\varepsilon_{\qq} = (\hbar
\mathbf{q})^2/2m_b$ the free particle energy, $\hbar\omega_\qq=
\sqrt{\varepsilon_{\qq} (\varepsilon_{\mathbf{q}}+2g n_0)}$ the
Bogoliubov dispersion relation, and $f_j(\qq)=\Omega^{-1/2}\int
\dd \mathbf{r} |\eta_j(\br)|^2\exp(\ii \qq\cdot\br)$, with
$\Omega$ the quantization volume. We note that, for $|\qq|\ll
1/\xi$, we have $M_{j,\qq}\propto f_j(\qq)/\sqrt{|\qq|}$, whereas
for $|\qq|\gg 1/\xi$, one obtains $M_{j,\qq}\propto
f_j(\qq)/\qq^2$.

The Hamiltonian $\hat{H}_\mathrm{eff}$ describes the dynamics of
\emph{hopping} polarons according to an extended Hubbard model
\cite{Lewenstein-2006}, provided that $c\gg aJ/\hbar$, with
$c\sim\sqrt{g n_0/m_b}$ the phonon velocity and $a$ the lattice
spacing. We gain qualitative insight into the dependence of
$V_{i,j}$ and the constant polaronic level shift $E_\nu \equiv
E_p$ on the system parameters by considering a one-dimensional
quasi-BEC in the thermodynamic limit. We assume a sufficiently
deep lattice to approximate the Wannier functions by Gaussians of
width $\sigma\ll\xi$, and find $V_{i,j}=(\kappa^2/\xi
g)\,\ee^{-2|i-j|a/\xi}$ and $E_p = \kappa^2/2\xi g$. We note that
the interaction between impurities is always attractive. More
importantly, for realistic experimental parameters, $\xi\sim a$,
and hence the off-site terms $V_{j,j+1}\propto\ee^{-2a/\xi}$ are
non-negligible. This interaction potential is a direct consequence
of the local deformation of the BEC around each impurity, as shown
in Fig.~\ref{Fig:Scheme}. For a set of static impurities at
positions $x_j=a j$, the overall deformation of the BEC density to
order $\kappa$ is given by $n(x) = n_0 +
\sum_j\langle\an{X}_j\cre{b}_\qq\an{b}_\qq\cre{X}_j\rangle = n_0
-(\kappa/g\xi)\sum_j\ee^{-2|x-x_j|/\xi}$.

%%%%%%%%%%%%%%%%%%%%%%%%%%%%%%%%%%%%%%%%%%%%%%%%%%%%%%%%%%%%%%%%%%%%
%%%%%%%%%%%%%%%%%%%%%%%%%%     Transport    %%%%%%%%%%%%%%%%%%%%%%%%
%%%%%%%%%%%%%%%%%%%%%%%%%%%%%%%%%%%%%%%%%%%%%%%%%%%%%%%%%%%%%%%%%%%%

\textit{Coherent and diffusive transport.---}We first consider
coherent hopping of polarons at small BEC temperatures $k_B T \ll
E_p$, where incoherent phonon scattering is highly suppressed.
Provided that $\zeta=J/E_p\ll 1$, we can apply the so-called
strong-coupling theory \cite{Alexandrov-PRB-1986}, and treat the
hopping term in Eq.~(\ref{htrans}) as a perturbation. Including
terms of first order in $\zeta$, we obtain the impurity
Hamiltonian

\begin{eqnarray}
\label{Eq:H1}\nonumber \hat H^{(1)}&=&-\tilde J \sum_{\langle i,j
\rangle}\cre{a}_i\an{a}_j+\frac{1}{2}\tilde{U}\sum_j\hat n_j(\hat n_j-1)+  \tilde{\mu}\sum_j\hat n_j\\
&&-\frac{1}{2}\sum_{i \neq j}V_{i,j}\,\hat n_i \hat n_{j}\,,
\end{eqnarray}
with $\tilde{\mu} = \mu+\kappa n_0-E_p$, $\tilde{U}=U-2E_p$, and
$\tilde J = J\,\langle\!\langle \hat X_i^\dagger \hat
X_j\rangle\!\rangle$. Here
$\langle\!\langle\,\cdot\,\rangle\!\rangle$ denotes the average
over the thermal phonon distribution and $i,j$ are nearest
neighbors. We find $\langle\!\langle \hat X_i^\dagger \hat
X_j\rangle\!\rangle = \exp\big\{-\sum_{\qq \neq 0}
|M_{0,\mathbf{q}}|^2 [1-\cos(\mathbf{q} \cdot
\mathbf{a})](2N_\mathbf{q}+1)\big\}$, with $\mathbf{a}$ the
position vector connecting two nearest neighbor sites and
$N_\mathbf{q}=(\ee^{\hbar\omega_\mathbf{q}/k_B T}-1)^{-1}$. Thus,
the hopping bandwidth decreases exponentially with increasing
coupling constant $\kappa$ and temperature $T$.

%%%%%%%%%%%%%%%%%%%%%%%%%%%%%%%%%%%%%%%%%%%%%%%%%%%%%%%%%%%%%%%%%%%%
%%%%%%%%%%%%%%%%%%%%%%%%%%   Diffusion  %%%%%%%%%%%%%%%%%%%%%%%%%%%%
%%%%%%%%%%%%%%%%%%%%%%%%%%%%%%%%%%%%%%%%%%%%%%%%%%%%%%%%%%%%%%%%%%%%

\begin{figure}
  \includegraphics[width=8cm]{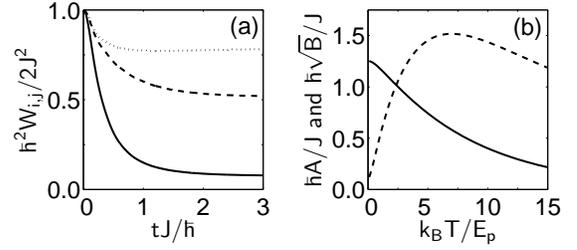}
\caption{(a) Memory function $\hbar^2 W_{i,j}(t)/(2 J^2)$ versus
time for $k_BT=0$ (dotted line), $2\,E_p$ (dashed line) and
$10\,E_p$ (full line). It drops off rapidly for $k_BT \gg E_p$,
indicating the dominance of incoherent hopping. (b) Coefficients
$\hbar A/J$ (dashed line) and $\hbar\sqrt{B}/J$ (full line) versus
temperature, obtained from the numerical solution of the GME in
Eq.~(\ref{GME}) for an evolution time $\tau=10\hbar/J
\approx0.8\times10^{-3}$s. The condition $A=\sqrt{B}$ is satisfied
at $k_BT/E_p\approx 2.3$, with $E_p/k_B\approx 11$nK. The lattice
(wavelength $\lambda = 790$nm) contains a single
${}^{41}\mathrm{K}$ atom with $J = 2.45\times 10^{-2}E_R$,
$\kappa/E_R \lambda = 2.3\times10^{-2}$, the recoil energy $E_R =
(2 \pi \hbar)^2/2m_a \lambda^2$, and $m_a$ the mass of the lattice
atom. The BEC consists of ${}^{87}$Rb atoms with $n_0 = 5\times
10^6\mathrm{m}^{-1}$ and $g/E_R\lambda =8.9\times10^{-3}$. }
  \label{Fig:Diff}
\end{figure}

At high temperatures $E_p\ll k_B T\ll k_B T_c$ (with $T_c$ the
critical temperature of the BEC) inelastic scattering, in which
phonons are emitted and absorbed, becomes dominant, and thus the
transport of atoms through the lattice changes from being purely
coherent to incoherent. We investigate this crossover by deriving
a generalized master equation (GME) for a single particle starting
from $\hat{H}_\mathrm{eff}$ in Eq.~(\ref{hlf}). Using the
Nakajima-Zwanzig projection method \cite{Breuer-2002}, we find
that the occupation probabilities $P_l(t)$ at site $l$ and time
$t$ evolve according to the GME \cite{Kenkre-PRB-1975}
\begin{equation}\label{GME}
 \frac{\partial P_i(t)}{\partial t}= \int_0^t \! \dd s \,
 \sum_{j}
\big[W_{i,j}(s)P_j(t-s) - W_{j,i}(s)P_i(t-s)\big]\,,
\end{equation}
where the effect of the phonon bath is encoded in the memory
functions $W_{i,j}(s)$, which are symmetric in $(i,j)$. To first
order in $\zeta$, thus keeping only nearest-neighbor correlations,
we find
\begin{eqnarray}\label{memory}\nonumber
    W_{i,j}&&\hspace{-10pt}(s)=2\bigg(\frac{J}{\hbar}\bigg)^2\Re\bigg[\exp\Big\{2\sum_{\qq \neq 0}
    |M_{0,\mathbf{q}}|^2
    [1-\cos(\mathbf{q} \cdot
    \mathbf{a})] \\
    &&\times[(N_\mathbf{q}+1)(\ee^{\ii\omega_{\mathbf{q}}
    s}-1) + N_\mathbf{q}(\ee^{-\ii\omega_{\mathbf{q}}
    s}-1)]\Big\}\bigg]\,,
\end{eqnarray}
and $W_{i,j}(s) = 0$ if $i$ and $j$ are beyond nearest neighbors.
The nontrivial part of $W_{i,j}(s)$ takes the values
$2(J/\hbar)^2$ at $s=0$ and $2(\tilde{J}/\hbar)^2$ in the limit
$s\rightarrow\infty$. In the regime $k_B T \ll E_p$, we have
$\tilde{J}\sim J$, and the memory function $W_{i,j}(s)$ is well
approximated by $2(\tilde{J}/\hbar)^2\Theta(s)$ [with $\Theta(s)$
the Heaviside step function], which describes purely coherent
hopping in agreement with $\hat{H}^{(1)}$. For $E_p \ll k_B T \ll
k_B T_c$, we observe that $\tilde{J}\ll J$, and $W_{i,j}(s)$ drops
off sufficiently fast for the Markov approximation to be valid, as
illustrated in Fig.~\ref{Fig:Diff}(a). In this case one can
replace $P_l(t-s)$ by $P_l(t)$ in Eq.~(\ref{GME}) and after
integration over $s$ the GME reduces to a standard Pauli master
equation $\partial_{t}P_i(t)=\sum_j w_{i,j}[P_j(t)-P_i(t)]$
describing purely incoherent hopping. The hopping rate is of the
form $w_{i,j} \sim J^2\exp(-E_a/k_B T)/(\hbar\sqrt{k_B T E_a})$
\cite{Holstein-Ann-1959,Mahan-2000}, where $E_a \sim E_p$ is the
activation energy.

The temperature-dependent crossover from coherent to diffusive
hopping  in a quasi-one-dimensional (1D) system is apparent in the
evolution for a time $\tau\sim 1/w_{i,j}$ of a particle initially
localized at lattice site $j=0$. The mean-squared displacement of
the lattice atom, $\overline{l^2}(t) = \sum_l l^2 P_l(t)$, can be
decomposed as $\overline{l^2}(t) = At+Bt^2$ into incoherent and
coherent contributions, characterized by the coefficients $A$ and
$B$, respectively. The crossover takes place when incoherent and
coherent contributions to $\overline{l^2}(\tau)$ are comparable,
i.e., $\sqrt{B} = A$. Fig.~\ref{Fig:Diff}(b) shows $A$ and
$\sqrt{B}$ as functions of $T$, where $\overline{l^2}(t)$ was
obtained by numerically solving the GME using the memory function
$W_{i,j}$ in Eq.~(\ref{memory}). We find that, for a
${}^{41}$K-${}^{87}$Rb system \cite{PhysRevLett.89.190404} under
standard experimental conditions, the crossover takes place well
below the critical temperature of the BEC (see caption of
Fig.~\ref{Fig:Diff}).

%%%%%%%%%%%%%%%%%%%%%%%%%%%%%%%%%%%%%%%%%%%%%%%%%%%%%%%%%%%%%%%%%%%%
%%%%%%%%%%%%%%%%%%%%%%%%%%%%%%%%%%%%%%%%%%%%%%%%%%%%%%%%%%%%%%%%%%%%

%%%%%%%%%%%%%%%%%%%%%%%%%%%%%%%%%%%%%%%%%%%%%%%%%%%%%%%%%%%%%%%%%%%%
%%%%%%%%%%%%%%%%%%%%%%%  Polaron Clusters %%%%%%%%%%%%%%%%%%%%%%%%%%
%%%%%%%%%%%%%%%%%%%%%%%%%%%%%%%%%%%%%%%%%%%%%%%%%%%%%%%%%%%%%%%%%%%%

\begin{figure}
  \includegraphics[width=7.75cm]{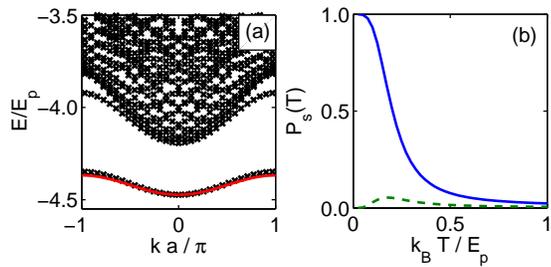}
  \caption{(Color online) (a) Energy spectrum of three polarons in a
  1D optical lattice. The solid line shows the lowest energy band $E_3(k)$
  characterizing off-site three-polaron clusters.
  The lattice (wavelength $\lambda = 790$nm, $M=31$ sites,
  periodic boundary conditions)
  contains$~^{133}\mathrm{Cs}$ atoms with $\tilde J = 7.5\times10^{-3}E_R$,
  $U = 50 E_p$, and $\kappa/E_R \lambda = 1.05\times10^{-1}$.
  The BEC consists of$~^{87}\mathrm{Rb}$ atoms with
  $n_0 = 5 \times 10^{6}\,\mathrm{m}^{-1}$ and $g/E_R \lambda =
  4.5\times10^{-2}$. (b) Probability of finding a three-polaron cluster (solid line)
  or a two-polaron cluster (dashed line) versus temperature.
  The parameters are $g/E_R \lambda = 6.5\times10^{-2}$,
  $\kappa/E_R \lambda = 1.32\times10^{-1}$, $U = 2.2 E_p$, $M=27$, and the rest as for (a).
  \label{Fig:TriPolaron}}
\end{figure}

\textit{Polaron Clusters.---}We now discuss the formation of
polaron clusters for $k_B T \lesssim E_p$, based on Hamiltonian
$\hat H^{(1)}$ in Eq.~(\ref{Eq:H1}), and assume that the bosonic
impurities are in thermal equilibrium with the BEC. At these
temperatures $V_{i,j}$, and $\tilde J$ are well approximated by
their $T=0$ values. We consider the limit $\tilde{U}\gg
V_{j,j+1}$, $\tilde{U}\gg \tilde J$, and adiabatically eliminate
configurations with multiply occupied sites. Keeping only nearest
neighbor interactions, we obtain approximate expressions for the
binding energy $E_b(s)\approx (s-1) V_{j,j+1}$ of a cluster of $s$
polarons located in adjacent sites and the lowest energy band
$E_k(s)\approx -E_b(s)-2\tilde{J}^s (V_{j,j+1})^{1-s}\cos(ka)$
\cite{Scott-PhysicaD-1994}, with $k$ the quasimomentum. This band
approximation is in good agreement with the results from exact
diagonalization of $\hat{H}^{(1)}$ using the full interaction
potential $V_{i,j}$, as shown for three polarons in
Fig.~\ref{Fig:TriPolaron}(a).

%%%%%%%%%%%%%%%%%%%%%%%%%%%%%%%%%%%%%%%%%%%%%%%%%%%%%%%%%%%%%%%%%%5
%%%%%%%%%%%%%%%%%%%%%%%%%%%%%%%%%%%%%%%%%%%%%%%%%%%%%%%%%%%%%%%%%%%%

This model predicts a decreasing average cluster size with
increasing temperature. For a small system with $N=3$ polarons, we
calculate the probability of finding a three-polaron cluster $P_3
= \sum_\nu \langle n_\nu \rangle$, where the sum is taken over all
states with energies $\varepsilon_\nu <E_g(2) + E_g(1)$, $E_g(N)$
is the ground state energy of an $N$-polaron cluster, and the
occupation probabilities are given by the Boltzmann law $\langle
n_\nu \rangle \propto \exp(-\varepsilon_\nu / k_B T)$.
Analogously, we determine the probability $P_2$ of finding a
two-polaron cluster or bipolaron. The results are shown in
Fig.~\ref{Fig:TriPolaron}(b). The probability of having a
three-polaron cluster goes down with increasing temperature and
for $T = E_p/k_B \approx 18\mathrm{nK}$ is essentially zero for
the parameters chosen. For this three-polaron cluster,
three-particle loss is negligible due to the on-site repulsion
$U$. Decreasing the value of $U$ gives a significantly increased
$P_2$ compared to the values shown in
Fig.~\ref{Fig:TriPolaron}(b).

\begin{figure}
  \includegraphics[width=7.75cm]{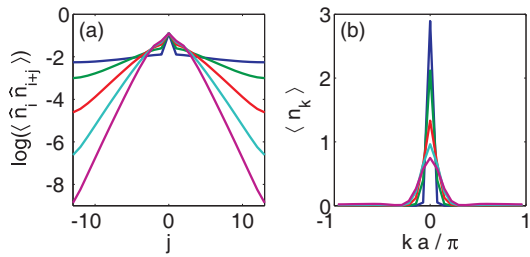}
  \caption{(Color online)
  (a) Density-density correlation in a system of three polarons indicating the
  formation of off-site three-polaron clusters.
  (b) Corresponding momentum distribution $\langle n_k \rangle$.
  $\kappa/E_R \lambda = \{\,4.0,\,6.1,\,8.1,\,10.1,\,12.1\,\}\times10^{-2}$
  where a higher value corresponds to a more localized
  correlation in (a) and a more spread distribution in (b).
  The other parameters are $g /E_R \lambda= 6.5\times10^{-2}$, $M=27$,
  $U = 3 E_p$, and
  the rest as in Fig.~\ref{Fig:TriPolaron}(a).
  \label{Fig:Clusters}}
\end{figure}

The clustering of polarons leads to their mutual exponential
localization.  This is illustrated by the density-density
correlations $\langle \hat n_{i} \hat n_{i+j} \rangle$ for the
three-polaron cluster in Fig.~\ref{Fig:Clusters}(a). With
increasing attractive interaction $V_{i,j}$, the mutual
localization gets stronger, leading to an increased broadening of
the momentum distribution, as shown in Fig.~\ref{Fig:Clusters}(b).
This allows polaron clusters to be identified in time of flight
experiments. We note that the transition from a superfluid to a
Mott insulator also leads to broadening of the momentum
distribution as, e.g., observed in
\cite{ref-exp-fermi-bose-short}. However, using Bragg spectroscopy
\cite{Winkler-nature-2006-short} would allow the unambiguous
distinction of boson clustering from this transition.

%%%%%%%%%%%%%%%%%%%%%%%%%%%%%%%%%%%%%%%%%%%%%%%%%%%%%%%%%%%%%%%%%%
%%%%%%%%%%%%%%%%%%%%%%%%%%%%%%%%%%%%%%%%%%%%%%%%%%%%%%%%%%%%%%%%%%%%

%%%%%%%%%%%%%%%%%%%%%%%%%%%%%%%%%%%%%%%%%%%%%%%%%%%%%%%%%%%%%%%%%%%%
%%%%%%%%%%%%%%%%%%%%%%%%%%  CONCLUSION  %%%%%%%%%%%%%%%%%%%%%%%%%%%%
%%%%%%%%%%%%%%%%%%%%%%%%%%%%%%%%%%%%%%%%%%%%%%%%%%%%%%%%%%%%%%%%%%%%

\textit{Conclusion.---}We have demonstrated that the dynamics of
bosonic impurities immersed in a BEC is accurately described in
terms of polarons. We found that spatial coherence is destroyed in
hopping processes at large BEC temperatures while the main effect
of the BEC at low temperatures is to reduce the coherent hopping
rate. Furthermore the phonons induce off-site interactions which
lead to the formation of stable clusters which are not affected by
loss due to inelastic collisions. Using the techniques introduced
in this paper qualitatively similar phenomena can also be shown to
occur for fermionic impurities. In either case these effects can
be controlled by external parameters and lie within the reach of
current experimental techniques.

\acknowledgments{\textit{Acknowledgements.---}M.B. thanks Danail
Obreschkow for fruitful discussions. This work was supported by
the U.K. EPSRC through QIP IRC (Grant No. GR/S82176/01) and
project No. EP/C51933/1, the EU through the STREP project OLAQUI,
the Berrow Scholarship (M.B.), and the Keble Association (A.K.).}

\vspace{-0.35cm}

\bibliography{short}

\end{document}